\documentclass[12pt,preprint]{aastex}

\slugcomment{To appear in ApJL}

\newcommand\refeq[1]{eq.~(\ref{eq:#1})}
\newcommand\reffig[1]{Fig.~\ref{fig:#1}}

\begin{document}

\title{Rethinking Lensing and $\Lambda$}
\author{Charles R. Keeton\altaffilmark{1}}
\affil{
  Astronomy and Astrophysics Department,
  University of Chicago, \\
  5640 S. Ellis Ave.,
  Chicago, IL 60637
}
\altaffiltext{1}{Hubble Fellow}

\begin{abstract}
Strong gravitational lensing has traditionally been one of the few
phenomena said to oppose a large cosmological constant; many analyses
of lens statistics have given upper limits on $\Omega_\Lambda$ that
are marginally inconsistent with the concordance cosmology. Those
conclusions were based on models where the predicted number counts
of galaxies at moderate redshifts ($z \sim 0.5$--1) increased
significantly with $\Omega_\Lambda$.  I argue that the models
should now be calibrated by counts of distant galaxies. When this
is done lens statistics lose most of their sensitivity to the
cosmological constant.
\end{abstract}


\section{Introduction}

Popular opinion seems to have settled on a ``concordance''
cosmology dominated by dark energy. The cosmic microwave background
indicates a flat geometry (e.g., de Bernardis et al.\ 2000; Hanany
et al.\ 2000; Pryke et al.\ 2002), cluster mass-to-light ratios
indicate a low matter density (e.g., Carlberg, Yee \& Ellingson
1997; Bahcall et al.\ 2000), and type Ia supernovae indicate cosmic
acceleration (e.g., Riess et al.\ 1998; Perlmutter et al.\ 1999).
The popular cosmology has matter content $\Omega_M \approx
0.30$--0.35 and dark energy $\Omega_X \approx 0.65$--0.70 such that
$\Omega_M + \Omega_X = 1$. For simplicity I assume that the dark
energy corresponds to a cosmological constant $\Lambda$, although
my analysis could easily be extended to quintessence models (e.g.,
Waga \& Miceli 1999).

One phenomenon that has traditionally stood out from the
concordance model is strong gravitational lensing. The statistics
of strong lenses are sensitive to the cosmological parameters via
the cosmological volume element (e.g., Turner 1990; Fukugita,
Futamase \& Kasai 1990), and analyses of the data have yielded
upper limits on the dark energy at the level of
$\Omega_\Lambda < 0.66$ at 95\% confidence (e.g., Kochanek 1996;
Falco, Kochanek \& Mu\~noz 1998). There are small systematic
uncertainties in the upper limit due to assumptions about the lens
sample and the amount of dust extinction in lens galaxies (e.g.,
Falco et al.\ 1998; Helbig et al.\ 1999; Waga \& Miceli 1999;
Cooray, Quashnock \& Miller 1999). Larger systematic effects arise
from uncertainties in the local luminosity function of galaxies.
Chiba \& Yoshii (1999) argue that by adopting luminosity functions
from different surveys they can relax the upper limit on
$\Omega_\Lambda$ and even find models that favor values in the
range $\Omega_\Lambda \sim 0.5$--0.8. Kochanek et al.\ (1998)
respond by acknowledging the systematic uncertainties but defending
their choice of luminosity function as the one that is most
consistent with the observed luminosities of lens galaxies. This
controversy will soon be resolved by new measurements of the
luminosity function from the SDSS and 2dF surveys that appear to
eliminate most of the traditional problems (Blanton et al.\ 2001;
Cross et al.\ 2001).

While debating the details, the previous studies agreed on the idea
that raising $\Omega_\Lambda$ dramatically increases the expected
number of lenses; they agreed on the trend and only contested the
zero point. In this {\it Letter\/} I question the trend itself,
based on number counts of distant galaxies. I argue that the lens
statistics models used to obtain upper limits on $\Omega_\Lambda$
are inconsistent with galaxy counts at $z \sim 0.5$ --- not with
any particular value of the counts, but with the general idea that
they can be measured and used as constraints on the models. I modify
the models to be calibrated by galaxy counts and consider how the
new models depend on $\Omega_\Lambda$. For simplicity I consider
only flat cosmologies.

\section{Fundamentals}

The optical depth for lensing of sources at redshift $z_s$ is
\begin{equation}
  \tau(z_s) = \int_{0}^{z_s} dz_l\,D(z_l)^2\,\frac{dD}{dz_l}
    \int dM\,\frac{dn}{dM}\, A(M,z_l,z_s)\,,
\end{equation}
where $dn/dM$ is the mass function of deflectors, $A(M,z_l,z_s)$ is
the cross section for lensing by a deflector of mass $M$ at
redshift $z_l$, and $D(z)$ is the comoving distance. When analyzing
real lens data the optical depth must be modified to account for
magnification bias, or the fact that lensing magnification
effectively changes a survey's flux limit (e.g., Turner, Ostriker
\& Gott 1984). Magnification bias is largely insensitive to
cosmological parameters, so for our purposes the simple optical
depth is adequate.

Nearly all arcsecond-scale lenses are produced by detectable
galaxies, so many analyses of lens statistics model the deflector
population with observed galaxy populations. The luminosity
function of galaxies is often parameterized as a Schechter function
with the form
\begin{equation}
  \frac{dn}{dL} = \frac{n_*}{L_*} \left(\frac{L}{L_*}\right)^\alpha
    e^{-L/L_*} ,
\end{equation}
where $L_*$ is a characteristic luminosity, $n_*$ is a
characteristic comoving number density, and $\alpha$ is the power
law slope at the faint end. Luminosity must be converted to mass
for the lensing analyses, and this is usually done with the
empirical Faber--Jackson relation
$L/L_* = (\sigma/\sigma_*)^\gamma$ where $\sigma$ is the velocity
dispersion.

Luminosity naturally changes with redshift due to passive
evolution: stellar populations fade with time as stars die out. The
number density may change with time due to mergers. The number
evolution can be parameterized in various ways, but Lin et al.\
(1999) propose the form
\begin{equation}
  n_*(z) = n_*(0)\,10^{0.4 P z} , \label{eq:nstar}
\end{equation}
so that the number evolution parameter $P$ can be combined with
another parameter representing luminosity evolution expressed in
observers' units of magnitudes. I adopt this form since the exact
parameterization is not very important for my analysis. Other than
luminosity and number evolution, we may assume for simplicity that
$\alpha$, $\gamma$, and $\sigma_*$ are all independent of redshift
(see Lin et al.\ 1999).

Lens galaxies are usually approximated as isothermal spheres, which
is consistent with lens data, galaxy dynamics, and X-ray ellipticals
(e.g., Fabbiano 1989; Rix et al.\ 1997; Treu \& Koopmans 2002). The
lensing optical depth can then be written as (e.g., Kochanek 1993)
\begin{eqnarray}
  \tau(z_s) &=& \tau_* \int_{0}^{z_s} dz_l\ \frac{dD_{ol}}{dz_l}\,
    \left[\frac{ D_{ol} D_{ls} }{ D_{os} }\right]^2
    \frac{ n_*(z_l) }{ n_*(0) }\,, \label{eq:tau1a} \\
  \tau_* &=& 16\pi^3 r_H^3 n_*(0) \left(\frac{\sigma_*}{c}\right)^4
    \Gamma\left[ 1 + \alpha + \frac{4}{\gamma} \right] ,
\end{eqnarray}
where $D_{ol}$, $D_{os}$, and $D_{ls}$ are comoving distances between
the observer, lens, and source, written here in units of the Hubble
distance $r_H = c/H_0$.

\section{Results}

\begin{figure}
\plotone{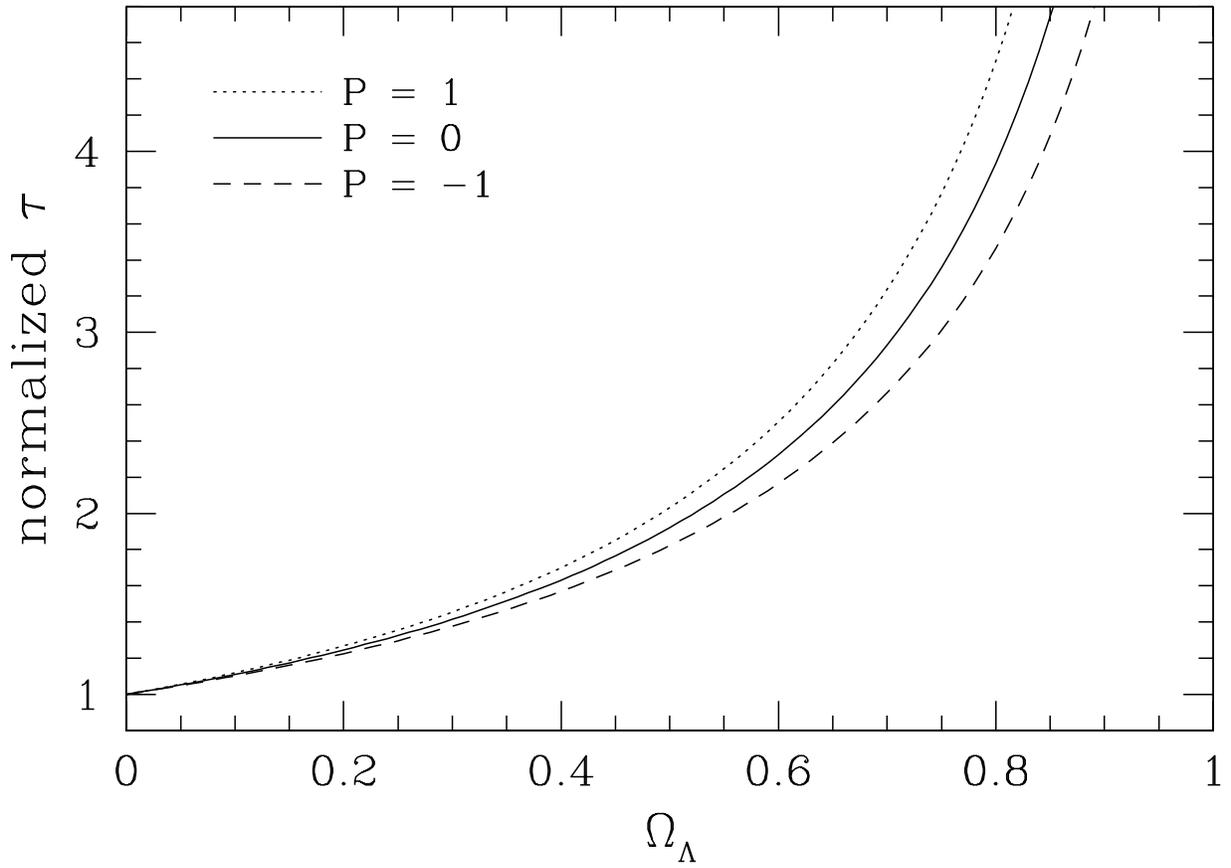}
\caption{
The lensing optical depth as a function of $\Omega_\Lambda$, for a
source redshift $z_s=2$. The optical depth is normalized by its
value at $\Omega_\Lambda=0$. The number evolution parameter $P$ is
assumed to be the same for all cosmologies.
}\label{fig:tau1}
\end{figure}

If there is no number evolution ($P=0$), then \refeq{tau1a} yields
the standard result for the lensing optical depth (e.g., Kochanek
1993),
\begin{equation}
  \tau(z_s) = {\tau_* \over 30}\, D_{os}^3\,, \label{eq:tau1b}
\end{equation}
where the optical depth is proportional to the cosmological volume
$D_{os}^3$ and hence very sensitive to $\Omega_\Lambda$, as shown
in \reffig{tau1}. If we allow for some number evolution but assume
that $n_*(z)$ is identical for all cosmologies (as done by Jain et
al.\ 2000), the optical depth curves shift slightly but retain the
same qualitative behavior. In other words, if we assume the same
number density (whether constant or not) for all cosmologies then
we recover the standard result that the number of lenses increases
significantly as $\Omega_\Lambda$ increases, with little
sensitivity to the actual amount of number evolution. The increase
is large enough that previous analyses of lens statistics were able
to obtain interesting upper limits on $\Omega_\Lambda$ even from
small lens samples.

\begin{figure}
\plotone{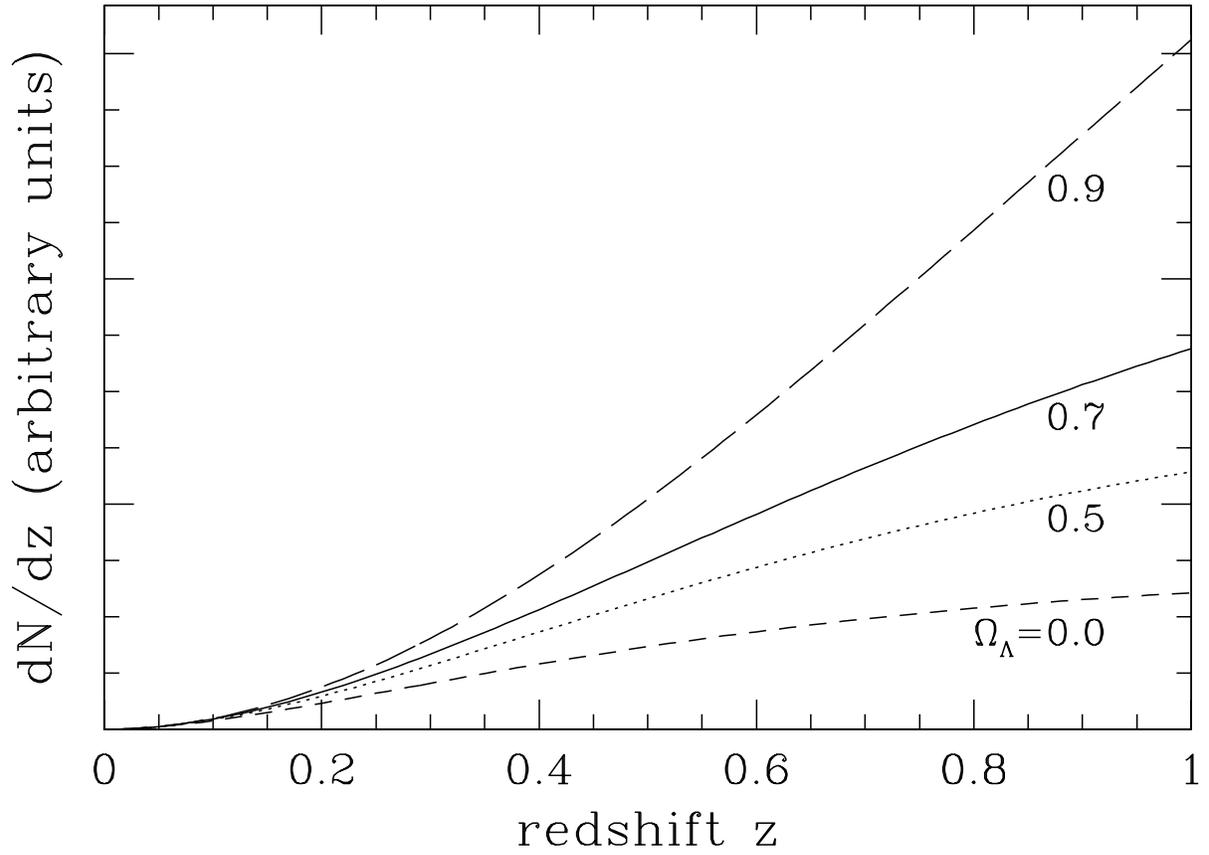}
\caption{
Galaxy number counts for the $P=0$ models from \reffig{tau1}.
Different line types indicate different values of $\Omega_\Lambda$.
}\label{fig:counts1}
\end{figure}

Although the focus of the models is the lensing optical depth, a
corollary prediction is the number of galaxies per unit redshift
per unit area on the sky,\footnote{For simplicity this analysis
omits luminosity cuts, but it would be straightforward to add them.}
\begin{equation}
  {dN \over dz} = n_*(z)\,D(z)^2\,{dD \over dz}\,
    \Gamma(1+\alpha)\,.
\end{equation}
\reffig{counts1} shows the predicted number counts for the models
from \reffig{tau1}. If $n_*(z)$ is the same in all cosmologies then
number counts of galaxies at moderate or high redshifts are very
sensitive to $\Omega_\Lambda$.  For example, at $z=0.5$ models with
$\Omega_\Lambda=0$ and 0.7 differ by a factor of two. This is the
basis of the classic idea to use galaxy number counts to constrain
the cosmology (see Sandage 1997; Driver 2002). Lens statistics boil
down in some sense to galaxy counts, where the objects being counted
are lens galaxies (mostly massive ellipticals at $z \sim 0.3$--1).

Galaxy counts at moderate redshifts ($z \sim 0.5$--1) can now be
measured directly (e.g., Lin et al.\ 1999; Davis et al.\ 2001;
Eisenstein et al.\ 2001), so we can imagine using them to test the
models. Let us consider not any particular value for the counts,
but just the general idea that they can be measured. If the counts
at $z \sim 0.5$--1 are measured then at most one of the curves in
\reffig{counts1} can be correct. What happens if instead we insist
that all models agree with some observed galaxy counts $dN/dz$? In
an idealized situation where number counts could be measured well
at all redshifts, the optical depth would be
\begin{equation}
  \tau(z_s) = 16\pi^3 \left({\sigma_* \over c}\right)^4
    { \Gamma(1+\alpha+4/\gamma) \over \Gamma(1+\alpha) }
    \int_{0}^{z_s} dz_l\, \left({D_{ls} \over D_{os}}\right)^2 {dN \over dz}\ .
    \label{eq:tau1c}
\end{equation}
In other words, the optical depth would simply be an integral over
the observed number counts, weighted by the distance ratio
$D_{ls}/D_{os}$. The direct dependence on the cosmological volume
$D_{os}^3$ would disappear, leaving only a weak dependence on
cosmology through the distance ratio.

In practice the number counts will be measured well for some
finite range of redshifts. Given counts $dN/dz$ at redshift
$z_{\rm obs}$ and parameterizing the evolution as in \refeq{nstar},
we would infer a number evolution parameter\footnote{If counts
were known over a range of redshifts, this expression and \refeq{P}
would include appropriate averages over $z_{\rm obs}$.}
\begin{equation}
  P = \frac{2.5}{z}\,\log\left(
    \frac{ dN/dz }{ n_*(0) \Gamma(1+\alpha) D^2\,dD/dz }
    \right)\biggr|_{z_{\rm obs}} .
\end{equation}
Note that this value depends on the assumed cosmology through the
volume element $D^2\,dD/dz$. Assuming two different cosmologies
(``1'' and ``2'') would lead to two different values of $P$ that
are related by
\begin{equation}
  P_2 - P_1 = {2.5 \over z}\,\log\left({D_1^2\,dD_1/dz \over
    D_2^2\,dD_2/dz}\right)\biggr|_{z_{\rm obs}} . \label{eq:P}
\end{equation}
Thus, when models are calibrated by observed galaxy counts at
moderate redshifts, the function $n_*(z)$ cannot be the same for
all cosmologies. The number density is the number counts divided
by the cosmological volume element, so it necessarily depends on
the assumed cosmological parameters. By using the same $n_*(z)$
in all cosmologies, traditional models for lens statistics have
violated this condition.

\begin{figure}
\plotone{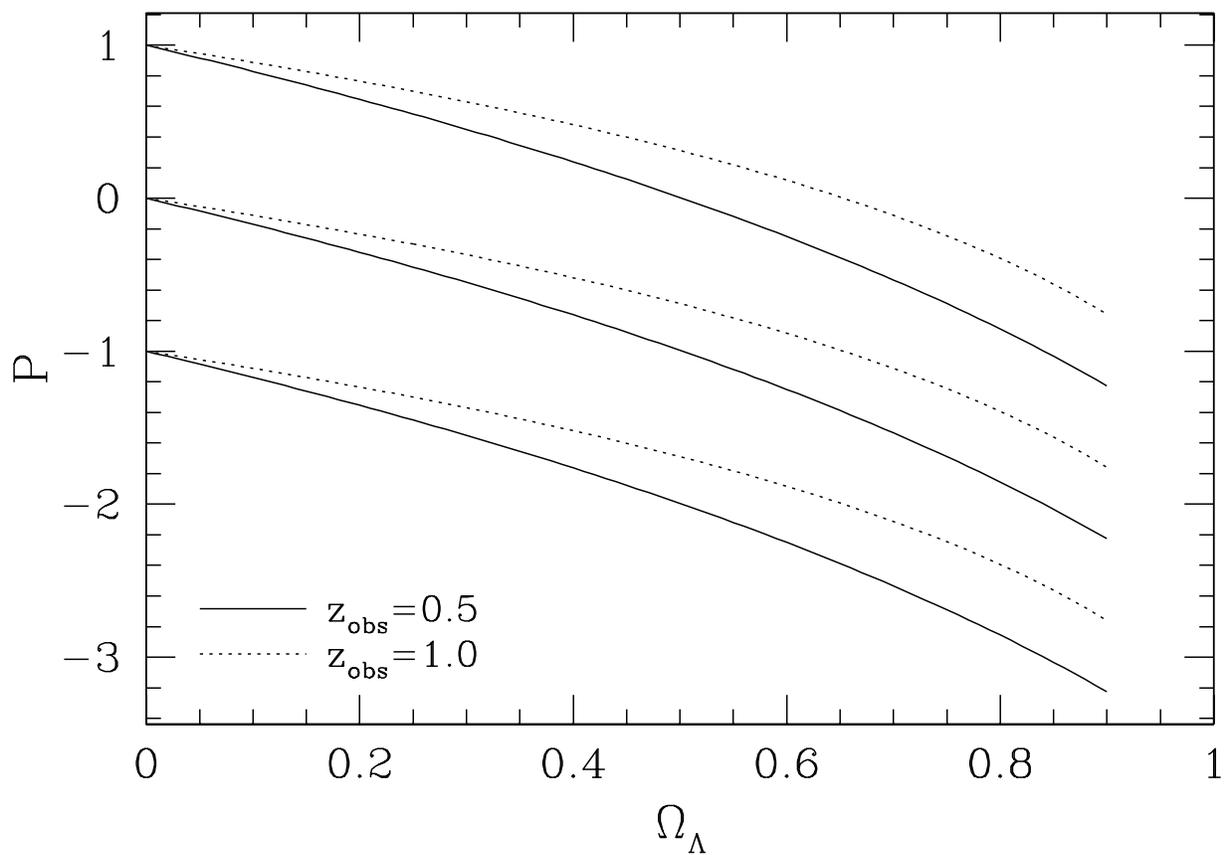}
\caption{
The change in the number evolution parameter $P$ with
$\Omega_\Lambda$ if the models are calibrated by galaxy number
counts at redshift $z_{\rm obs}=0.5$ (solid) or $z_{\rm obs}=1.0$
(dotted); see \refeq{P}. The curves correspond to $P_0=1$ (top), 0
(middle), and $-1$ (bottom).
}\label{fig:P}
\end{figure}

\begin{figure}
\plotone{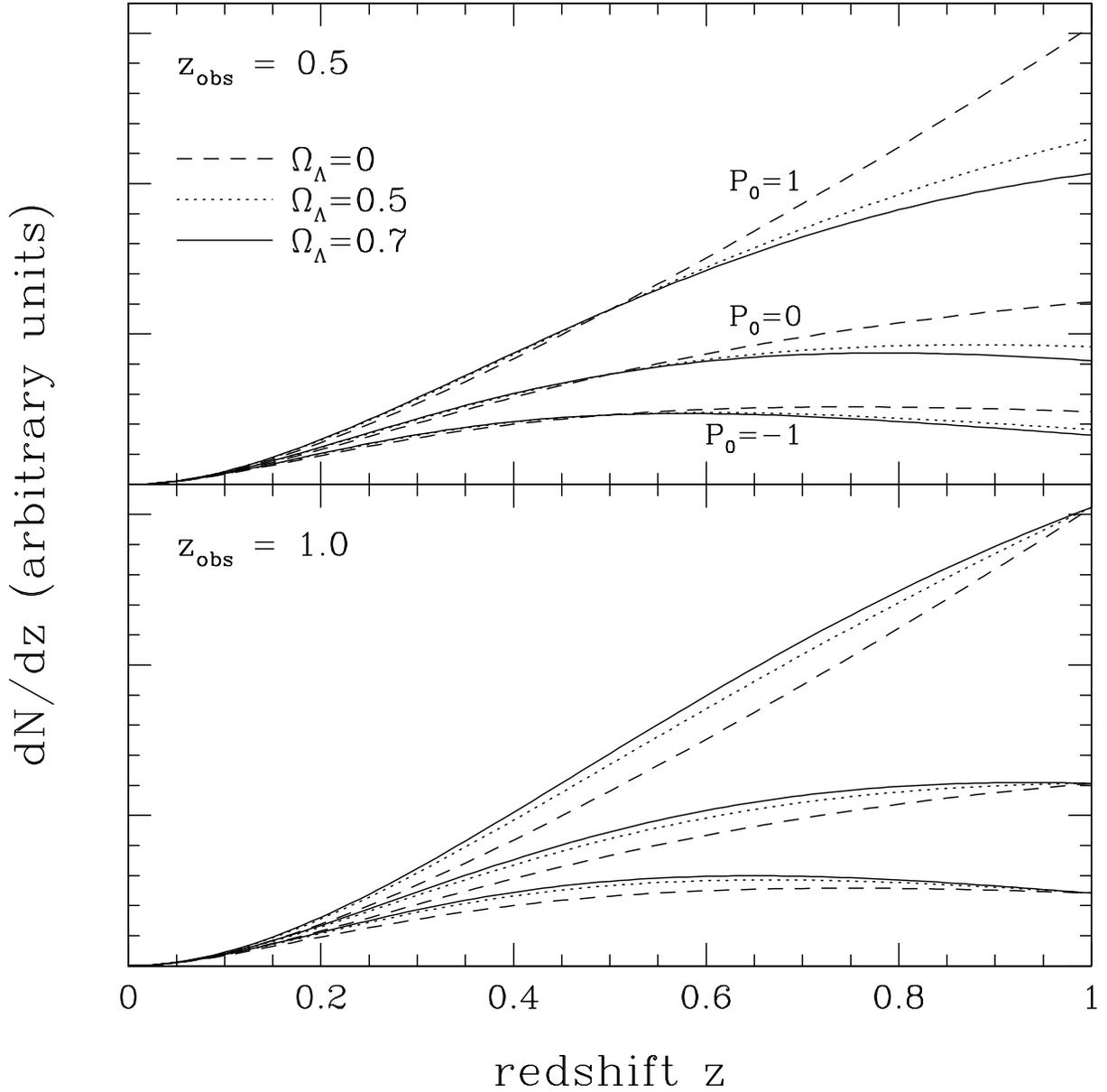}
\caption{
Galaxy number counts for new models calibrated by some observed
counts at $z_{\rm obs}=0.5$ (top) or $z_{\rm obs}=1.0$ (bottom).
Different line types indicate different values of $\Omega_\Lambda$.
The hypothetical measured counts correspond to $P_0=1$, 0, and $-1$
for the top, middle, and bottom set of curves in each panel. Note
that in the limit $z_{\rm obs} \to 0$ we would recover
\reffig{counts1}.
}\label{fig:counts2}
\end{figure}

Let us construct new models that do not violate it. Imagine that
from observed galaxy counts we infer a number evolution parameter
$P_0$ in a cosmology with $\Omega_\Lambda=0$. For other values of
$\Omega_\Lambda$ we can use \refeq{P} to compute the
self-consistent value of $P$, as in \reffig{P}. Models with
different values of $\Omega_\Lambda$ now agree on the predicted
number counts at $z_{\rm obs}$ (by construction) and are very
similar over a wide range of redshifts $z < z_{\rm obs}$, as shown
in \reffig{counts2}.\footnote{The various models could be made to
match over a wider range of redshift by introducing more parameters
to describe the evolution.} In other words, the new models are
consistent in that they agree on an observable quantity, galaxy
number counts, independent of $\Omega_\Lambda$.

\begin{figure}
\plotone{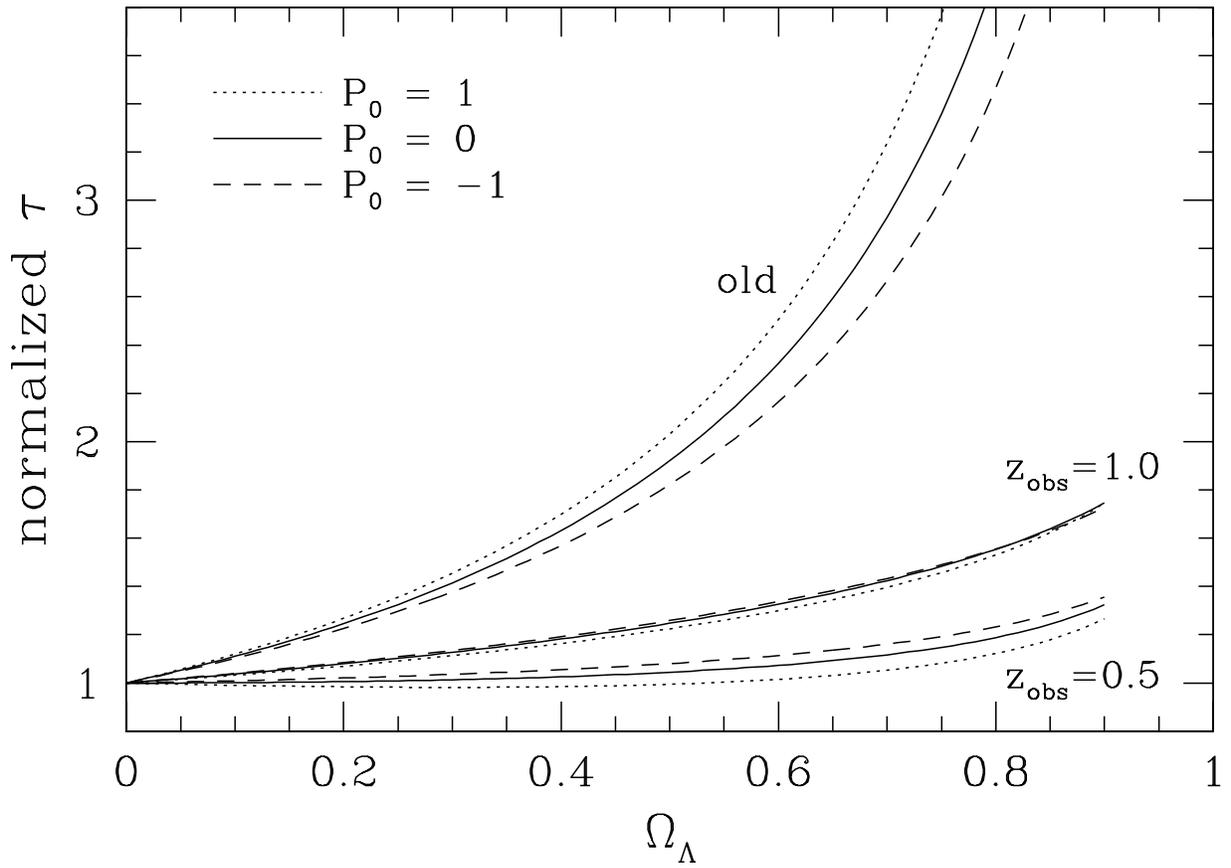}
\caption{
Similar to \reffig{tau1}, but with results for new models
calibrated by galaxy counts at redshift $z_{\rm obs}=0.5$ or $1.0$.
}\label{fig:tau2}
\end{figure}

\reffig{tau2} shows the lensing optical depth as a function of
$\Omega_\Lambda$ for these new models. The difference from the old
models is striking: the dependence of the optical depth on
$\Omega_\Lambda$ is much smaller than before. In the old models,
going from $\Omega_\Lambda=0$ to $\Omega_\Lambda=0.7$ increased
the number of lenses by a factor of 2.7--3.3 (depending on the
value of $P$), while in the new models the increase is only
40--43\% if the calibration is at $z_{\rm obs}=1.0$, or a mere
5--16\% if $z_{\rm obs}=0.5$. Models constrained by galaxy counts
at all redshifts as in \refeq{tau1c} (not shown) lie between the
$z_{\rm obs}=0.5$ and $z_{\rm obs}=1.0$ curves.

The reduced dependence on $\Omega_\Lambda$ is a direct result of
calibrating the models with observed counts of distant galaxies.
This qualitative conclusion does not depend on any particular
values I have assumed, but arises simply from the idea that
models for lens statistics can and should be constrained to agree
with measurable galaxy counts at redshifts relevant for lensing.
There is some residual sensitivity to cosmology through the
distance ratio $D_{ls}/D_{os}$ (see eq.~\ref{eq:tau1c}), but it
is much smaller than the volume effect and thus will be harder to
detect.

The prospects for using real galaxy counts to calibrate models
for lens statistics are good. The CNOC2 field galaxy redshift
survey already contains $\sim$5000 galaxies at $0.12<z<0.55$ (Lin
et al.\ 1999), the SDSS Luminous Red Galaxy sample will have
redshifts for $\sim$100,000 early-type galaxies out to $z \sim 0.5$
(Eisenstein et al.\ 2001), and the DEEP redshift survey will
include $\sim$60,000 galaxies at $z > 0.7$ (Davis et al.\ 2001).
The calibration of the models will be limited to some extent by
completeness and selection effects, but with large samples it
will be possible to make different cuts on the data to understand
those effects (see Blanton et al.\ 2001). The calibration will
also be limited by our understanding of the relationship between
luminosity and mass --- but that has always been true for models
of lens statistics. The best bet is probably to turn the problem
around: rather than trying to constrain $\Omega_\Lambda$, a joint
study of galaxy counts and lensing could attempt to distinguish
between evolution in luminosity and evolution in mass. That would
be perhaps the most interesting future application of lens
statistics.

\section{Conclusions}

The optical depth for lensing is basically proportional to the
number of galaxies on the sky at redshifts $z_l \sim 0.3$--1.
Traditionally that number was not well known, so models for lens
statistics adopted a number density of galaxies and multiplied by
the cosmological volume to get the number. With the assumed number
density held fixed, the expected number of lenses was proportional
to the cosmological volume and hence very sensitive to
$\Omega_\Lambda$.

Number counts of distant galaxies can now be measured directly, and
they are inconsistent with the idea that the number density is
independent of $\Omega_\Lambda$. This general point holds whether
there is much or little redshift evolution in the galaxy
population. Constraining models for lens statistics to agree on
distant galaxy counts makes them far less sensitive to
$\Omega_\Lambda$. Whereas the old models saw a factor of 3 change
in the number of lenses between $\Omega_\Lambda=0$ and
$\Omega_\Lambda=0.7$, in the new models the change is
$\lesssim$10--40\%. Using lens statistics to constrain
$\Omega_\Lambda$ may still be possible, but will be difficult.

This paper has focused on models where the deflector population
is derived from observed galaxy populations, which I refer to as
phenomenology models. In an alternate class that I call theory
models, the deflector population is described with a mass function
from structure formation theory; the resulting models are sensitive
to cosmology not only through the volume element but also through
$\Omega_M$ and the growth of structure. In theory models the
predicted number of lenses {\it decreases\/} as $\Omega_\Lambda$
increases (for flat cosmologies; Porciani \& Madau 2000; Li \&
Ostriker 2002), which strongly disagrees with old phenomenology
models but is less different from my new models. In quintessence
models, phenomenology and theory models do agree that making the
equation of state more negative increases the predicted number of
lenses (Waga \& Miceli 1999; Sarbu, Rusin \& Ma 2001). The
differences between phenomenology and theory models clearly need
further study. Nevertheless, I would argue that the focus of lens
statistics should move away from constraining $\Omega_\Lambda$
and toward learning about the population of dark matter halos out
to $z \sim 1$.

\acknowledgements
I am very grateful to the referee, David Rusin, for important
criticisms and insightful comments. I also thank Huan Lin for
interesting discussions and Wayne Hu for comments on the
manuscript.
This work was supported by NASA through Hubble Fellowship grant
HST-HF-01141.01-A from the Space Telescope Science Institute, which
is operated by the Association of Universities for Research in
Astronomy, Inc., under NASA contract NAS5-26555.

\clearpage

\end{document}